%% file: main.tex
\documentclass[10pt,journal,compsoc]{IEEEtran}

\ifCLASSOPTIONcompsoc
  \usepackage[nocompress]{cite}
\else
  \usepackage{cite}
\fi

\usepackage[numbers]{natbib}
\usepackage{amsmath,amssymb,amsfonts}
\usepackage{algorithmic}
\usepackage{graphicx}

\ifCLASSINFOpdf
\else
\fi
\ifCLASSOPTIONcompsoc
 \usepackage[caption=false,font=footnotesize,labelfont=sf,textfont=sf]{subfig}
\else
 \usepackage[caption=false,font=footnotesize]{subfig}
\fi

\usepackage{url}

\usepackage{xcolor}

\begin{document}

\title{Pro or Anti? \\ A Social Influence Model of Online Stance Flipping}


\author{Lynnette Hui Xian Ng, Kathleen M. Carley \\
Carnegie Mellon University, Pittsburgh, USA
\IEEEcompsocitemizethanks{\IEEEcompsocthanksitem Lynnette Ng and Kathleen Carley are with the Center for Informed Democracy and Social Cybersecurity, Carnegie Mellon University, Pittsburgh, USA. \\
E-mail: \{huixiann,carley\}@andrew.cmu.edu}
}


\markboth{IEEE Transactions on Network Science and Engineering}%
{Ng and Carley: IEEE Transactions on Network Science and Engineering}



\IEEEtitleabstractindextext{%
\begin{abstract}
Social influence characterizes the change of an individual's stances in a complex social environment towards a topic.
Two factors often govern the influence of stances in an online social network: endogenous influences driven by an individual's innate beliefs through the agent's past stances and exogenous influences formed by social network influence between users. Both endogenous and exogenous influences offer important cues to user susceptibility, thereby enhancing the predictive performance on stance changes or flipping.
In this work, we propose a stance flipping prediction problem to identify Twitter agents that are susceptible to stance flipping towards the coronavirus vaccine (i.e., from pro-vaccine to anti-vaccine). 
Specifically, we design a social influence model where each agent has some fixed innate stance and a conviction of the stance that reflects the resistance to change; agents influence each other through the social network structure.
From data collected between April 2020 to May 2021, our model achieves 86\% accuracy in predicting agents that flip stances. 
Further analysis identifies that agents that flip stances have significantly more neighbors engaging in collective expression of the opposite stance, and 53.7\% of the agents that flip stances are bots and bot agents require lesser social influence to flip stances. 
\end{abstract}

\begin{IEEEkeywords}
coronavirus vaccine, social influence, stance analysis, social bot dynamics
\end{IEEEkeywords}}

\maketitle

\IEEEdisplaynontitleabstractindextext

%
\IEEEpeerreviewmaketitle

\ifCLASSOPTIONcompsoc
\IEEEraisesectionheading{\section{Introduction}\label{sec:introduction}}
\else
\section{Introduction}
\label{sec:introduction}
\fi

%
%
%
%
\IEEEPARstart{A}long line of work in social influence has studied how susceptible people are to modifying their stances towards a topic in response to the stances expressed around them. 
In a complex social environment like Twitter, social influence characterizes the changes of stance through a user's static conditions and the impact of the interpersonal influence from his surrounding social network \cite{doi:10.1080/0022250X.1990.9990069}. 

The 2020 coronavirus pandemic sent the world to a standstill. Researchers scrambled to develop a vaccine that would ease the pandemic. As the vaccine developed, public opinion formed two main polarizing camps, the pro-vaccine and anti-vaccine, as has always been the case in vaccination history.
These camps are fondly termed ``pro-vaxxers" and ``anti-vaxxers", characterized by their stance toward vaccination.
Polarization in vaccination discourse on social media platforms has been witnessed on Facebook regarding the MMR vaccine \cite{SCHMIDT20183606,doi:10.1177/2056305119865465} and on Twitter towards the HPV vaccine \cite{cossard2020falling}. 

When analyzing the vaccination debate online, two groups of users must be considered: bot and non-bot users. 
The susceptibility of non-bot users online is of concern because these users are real people who may make decisions based on their online influence, like the coronavirus vaccine hesitancy resulting from the spread of mRNA vaccine fear \cite{brumfiel_2021}.
The role of bots in shaping narratives in online vaccine discourse has been studied through observations of highly clustered anti-vaccine groups of bot users on Twitter \cite{doi:10.1177/2056305119865465}. Narrative spread on social media is a worrying trend because the network structure makes it difficult for health organizations to penetrate and spread legitimate vaccination news to non-bot users.  
While there are extensive observations on the influence social bots have on online conversations \cite{suarez2016influence,bessi2016social}, these works do not study how bot agents change their opinion to fit in their online network.

The seminal work by \citet{doi:10.1080/0022250X.1990.9990069} introduces a social influence model, where each agent has an expressed opinion that is a combination of the agent's innate opinion and external influence from their neighbors. 
In a Twitter social network, a Twitter user is an agent. The agent's innate opinion can be inferred by the previously posted tweets, and the neighbor influence can be estimated by the conversation network of agents. 
Agents change their expressed opinions to minimize social cost, or the disagreement between their expressed opinion and the expressed opinion of their neighbors, resulting in the observed stance flipping behavior \cite{gionis2013opinion}. 
Within the observation of the flipping of expressed opinion lies the observation of an agent's neighborhood. Key factors in this neighborhood are the presence of neighboring bots, are of the opposite stance as the actor and collectively express an opinion within a short period of time. 
We borrow ideas from studies on ``coordinated activity", which derives participating agents through the high-frequency use of similar actions, to identify the collective expression of an opinion. 

Changes in a person's stance can be detrimental to the public society, especially when the person has been convinced to do public harm. In the context of the coronavirus pandemic, an example of a detrimental stance change is the support of the anti-vaccination camp and proliferating anti-vaccine narratives, thus reducing vaccination uptake which further prolongs the pandemic situation. The study of this group can inform public organizations about vulnerable populations of individuals that are swayed toward participating in destructive actions. At the same time, stance changes for the better, i.e. changing allegiance to the pro-vaccination camp aids the public good. This group provides researchers with ideas for social influence persuasion techniques.
Despite the rich line of vaccination polarization studies and influence dynamics, to our knowledge, past studies have not focused on predicting vaccination stance flips with social influence factors. 
In this work, we aim to bridge the gap between analyzing the polarizing vaccination debate and identifying individuals susceptible to stance changes due to social influence. 
We hope our work can inform both public organizations and researchers alike in how the combination of the influence of a person's own convictions and his social network can impact his outlook towards an issue, identify such population groups and thus design appropriate messaging strategies.

\subsection{Our Contributions}  
In this work, we make the following key contributions: 
\begin{enumerate}
    \item \textit{Social influence model for prediction of an agent's susceptibility to stance flipping.} Based on the stance flipping behavioral analysis in the online coronavirus debate, we introduce a prediction problem that focuses on identifying agents who are likely to flip stances. We adapt the Friedkin-Johnsen social influence model \cite{doi:10.1080/0022250X.1990.9990069} to perform this prediction task. Our model successfully predicts 86\% of the agents whose stance flips in the context of the coronavirus vaccinations. 
    \item \textit{Analysis of linguistic and network factors towards stance flipping.} We study Twitter agents' endogenous linguistic factors and stance conviction from the agent's past tweets and the exogenous social network factors that influence stance changes.
    \item \textit{Comparison of the response of bots and non-bots to social influence.} We analyze the response of two key groups of agents -- bots and non-bots -- towards their surrounding social influence. Our results show that bots have less endogenous stance conviction and flip even with fewer neighbors with the opposite stance compared to non-bots. This furthers misinformation research as social bots are malleable agents that intentionally promote certain narratives.  
    \item \textit{Comparison of agent neighborhood between agents that flip and do not flip stances.} We perform a comparative statistical analysis of the neighborhood surrounding the agents that flip and do not flip stances. We observe that actors that flip stances are typically non-bots and have a higher proportion of neighbors that collectively express their stances. 
\end{enumerate}

\section{Related Work}
To the best of our knowledge, we are the first to study changes in vaccine stances under a social influence model. In this section, we review two major perspectives: (a) vaccine stance characterization and polarization and (b) the dynamics of social influence.

\subsection{Vaccine Stance Polarization}
A first step in analyzing vaccine stance polarization is annotating stances, a task commonly known as stance detection.
Stance detection is the task of ``automatically determining from text whether the author of the text is in favor of, against or neutral toward a proposition or target" \cite{mohammad2017stance}.  With vaccination stance, the two key stances are pro-vaccine (in favor of) and anti-vaccine (against), which creates a polarized discourse \cite{meyer2019vaccine,walter2020russian,tyagi2020divide}. 
A line of machine learning classifiers have been built to characterize tweets in terms of pro- or anti-vaccine \cite{darwish2020unsupervised,pavan2020twitter,kunneman2020monitoring}. 
Most of them construct a text representation of the tweet using n-grams \cite{skeppstedt2017automatic} or word embeddings, then enhance the feature set with other lexical features like emotions \cite{mahajan2019analyzing}, employ deep learning approaches to infer stances towards vaccination \cite{DANDREA2019209} or inferred stances through a network-based propagation algorithm \cite{kumar2020social}. 

With stance detection algorithms, we are then able to analyze polarizing stances on the vaccination debate on social media. \citet{cotfas2021longest} analyzed the change in stances over time for the coronavirus vaccine, while \citet{doi:10.1080/21645515.2021.1911216} change in stance behavior on Twitter before and after the arrival of the coronavirus pandemic.
Prior work has shown that both pro- and anti- vaccine groups exhibit different online behavior in terms of the vaccines discussed \cite{van2020communities,kang2017semantic}, reach and network structure \cite{Gargiulo2020}. 
In general, the two camps interact mostly in separate echo-chambers and consume information within their group but not across groups \cite{SCHMIDT20183606}.

\subsection{Social Influence}
The Friedkin-Johnsen social influence model \cite{doi:10.1080/0022250X.1990.9990069} has been commonly used in opinion studies \cite{amelkin2017polar,bindel2015bad}. It stems from Festinger's psychological theories of social comparison \cite{festinger1954theory} and cognitive dissonance \cite{festinger1957theory}, where agents act to relieve the psychological discomfort from their disagreement with others. In the model, each agent has an opinion strength which affects the agent's willingness to change opinion, and is influenced by the opinions of the surrounding neighbors. 
Prior work has adapted the model to identify a set of target agent whose positive opinion maximizes the overall positive opinion for a promotional campaign in a social network \cite{gionis2013opinion}; 
to model the filter bubble effect in online opinion polarization caused by an administrator that tries to reduce conflict between opinion groups
\cite{10.1145/3336191.3371825};
to model an equilibrium in a Reddit network that minimizes disagreement \cite{musco2018minimizing}
\citet{10.1145/2538508} further considered an inverse problem to investigate the effect of stubborn agents who influence others but do not change their opinion. 

Other social influence models for social networks involve the construction of an influence locality model to predict retweet behavior using personal attributes (e.g. number of followers/followees and reciprocal following relationships) \cite{zhang2015influenced}, using Markov state transitions to model neighborhood majority influence on an individual agent. 

Specifically on vaccination stance, \citet{10.1371/journal.pone.0060373} observed that an individual's conformity to social influence and initial level of susceptibility are crucial to an agent's expressed stance. \citet{germani2021anti} studied the anti-vaccination infodemic of the coronavirus vaccine, demonstrating how anti-vaccination supporters on Twitter are engaged in a much larger and more communicative community as compared to pro-vaccine supporters, which enabled the success of the anti-vaccination movement. 

Additionally, when studying online social media, we must consider the role of automated agents, or bots, in the conversation. Past work have found that there is a higher proportion of bots that spread anti-vaccine messages as compared to pro-vaccine messages \cite{doi:10.2105/AJPH.2018.304567} and have characterized personas of state-sponsored actors in the manipulation of polarizing health debates \cite{doi:10.1177/2056305119865465,doi:10.2105/AJPH.2019.305564}, and demonstrated that social probing activity with bots triggered emotional responses of individuals and led to a more polarized network \cite{aiello2012people}.

\subsection{Combining Social Influence and Stance Detection} 
In combining social influence and stance detection, \citet{hegselmann2002opinion} performed simulations around the dynamics of opinion formation, studying not only stance polarization but also the eventual consensus, which Friedkin termed ``norm formation" \cite{friedkin2001norm}.
Comparing stances on three time phases surrounding the polarizing BREXIT referendum debate, connected neighbors on Twitter tend to have similar opinions \cite{10.1007/978-3-319-65813-1_10}. Another study combined productivity measured by tweet posting with endorsement measured by retweeting to measure social influence, analyzing the two sides of the debate in terms of influential users \cite{Grcar2017}.

Stance changes have typically been observed in a debate setting, where linguistic factors and audience factors are combined to predict whether an undecided audience member would make a stand \cite{longpre-etal-2019-persuasion}. Political studies have also examined the ``flip-flopping" of stance in US electoral politics, branding the observation as an attribute of their conviction on issues brought up in presidential debates \cite{lempert2009flip} and other situations like gun-control \cite{bouton2014guns}. 

\subsection{Collective Expression of Agents}
The idea of collective expression of agents is derived from studies on ``coordinated activity" among social media agents, which have been shown to have an active role in artificially manipulating online information narratives \cite{ng2021coordinating,Weber2021,giglietto2020takes}.

Existing methods in detection of agents participating in coordinated activity relies on the identification of actions that are synchronized in time \cite{Pacheco_Hui_Torres-Lugo_Truong_Flammini_Menczer_2021,DBLP:journals/corr/abs-2105-07454}. These actions include using the same hashtags, the same @-mentions of other agents, or similar texts and images. In performing the same action, these agents are expressing the same narrative \cite{ng2021coordinating}, creating an online diaspora of collective expression. 

Several frameworks have been developed to identify coordinated action, from counts of the synchronicity of actions within a time period \cite{DBLP:journals/corr/abs-2105-07454}, to modeling temporal coincidence with temporal point processes and Gaussian Mixture Models \cite{10.1145/3447548.3467391}. 
These methods have characterized the collective expression of opinions of ``coordinated groups" in several high-key events: the promotion of anti-mask, anti-vaccine and anti-science conspiracy theories during the COVID19 pandemic \cite{10.1145/3447548.3467391}; the polarized anti- and pro- stance towards the 2019 Hong Kong Protest \cite{Pacheco_Hui_Torres-Lugo_Truong_Flammini_Menczer_2021}; the clusters of narratives during the 2021 Capitol Riots \cite{ng2021coordinating}.



\section{Social Influence Model for Vaccine Stance Predictions}
\label{sec:model}
In this section, we describe the social influence model we built. This model predicts whether an agent would flip his stance towards the coronavirus vaccine. 

Our problem statement is as follows: Given a Twitter agent, we use a social influence model to represent the agent's stance $Y$ towards the coronavirus vaccine (pro- or anti-vaccine) in terms of endogenous and exogenous variables $X$. 
Then we characterize how susceptible the agent is to social influence and predict whether the agent will flip his stance $Y$ from pro-vaccine to anti-vaccine or vice versa.

\subsection{The Social Influence Model}
We adapt the Friedkin-Johnsen social influence model \cite{doi:10.1080/0022250X.1990.9990069} and describe the formation of a stance towards the coronavirus vaccine (pro-vaccine or anti-vaccine) in terms of an agent's innate static variables and the interpersonal influences from other agents in the network. We describe an agent's stance in terms of variables and the process linking them. Specifically, an agent's stance towards the vaccine is dependent on the previous stances and linguistic cues of past tweets and the surrounding neighbor's information.

\textbf{Agent stance.} We define agent stances $Y$ with the following model: $Y_{agent} = XB$, in which $Y$ is an agent stance outcome score, $X$ is a $1 \times k$ matrix of scores on $k$ endogenous and exogenous variables of the agent and $B$ is a $kx1$ vector of coefficients giving the effects of each of the endogenous variables. In our study, we used agents' linguistic cues as endogenous variables and network values as exogenous variables. 

Since we are probably analyzing only but a subset of the variables that might affect an agent's stance, we partition the $X$ and $B$ matrices. That is, the equation is modified to represent observations and coefficients of a subset of variables as in $Y_{agent} = X_*B_*$.



\textbf{The Base Influence Model.}
The base influence model estimates the impact of an agent's past tweets and the influence from the agent's neighbors on the agent's stance. Neighbors are other agents that have made communication with the agent in focus. The opinions of these neighbors, or ``peers" in the social influence model, directly affect on an individual's opinions. On Twitter, this means a reply, retweet, or mention by either neighbor agent and agent in focus. 

\begin{figure*}[!tbp]
  \centering
  \subfloat[Illustration of the calculation of influence weight $w$ of each neighbor, taking in account their distance from the agent in focus.]{\includegraphics[width=0.4\textwidth,height=0.2\textheight]{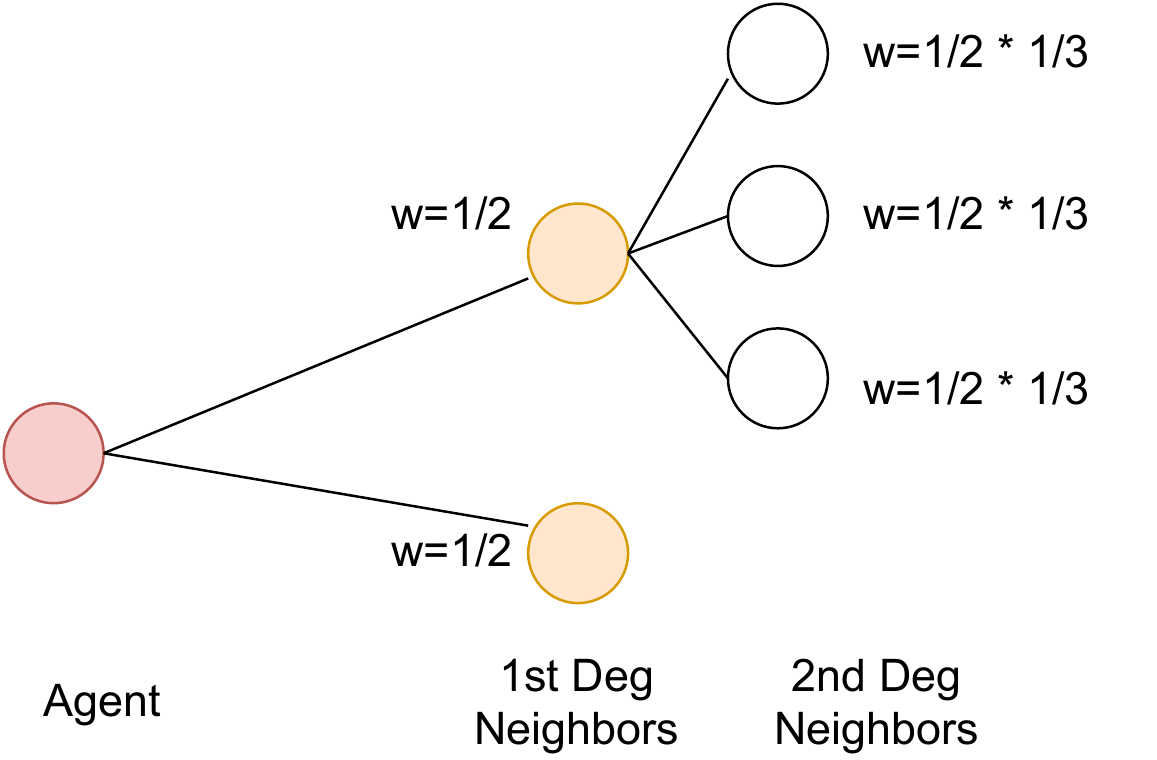}\label{fig:influenceweight}}
  \hfill
  \subfloat[Influence weight of each neighbor tends to 0 as the number of hops from the agent increases.]{\includegraphics[width=0.4\textwidth,height=0.25\textheight]{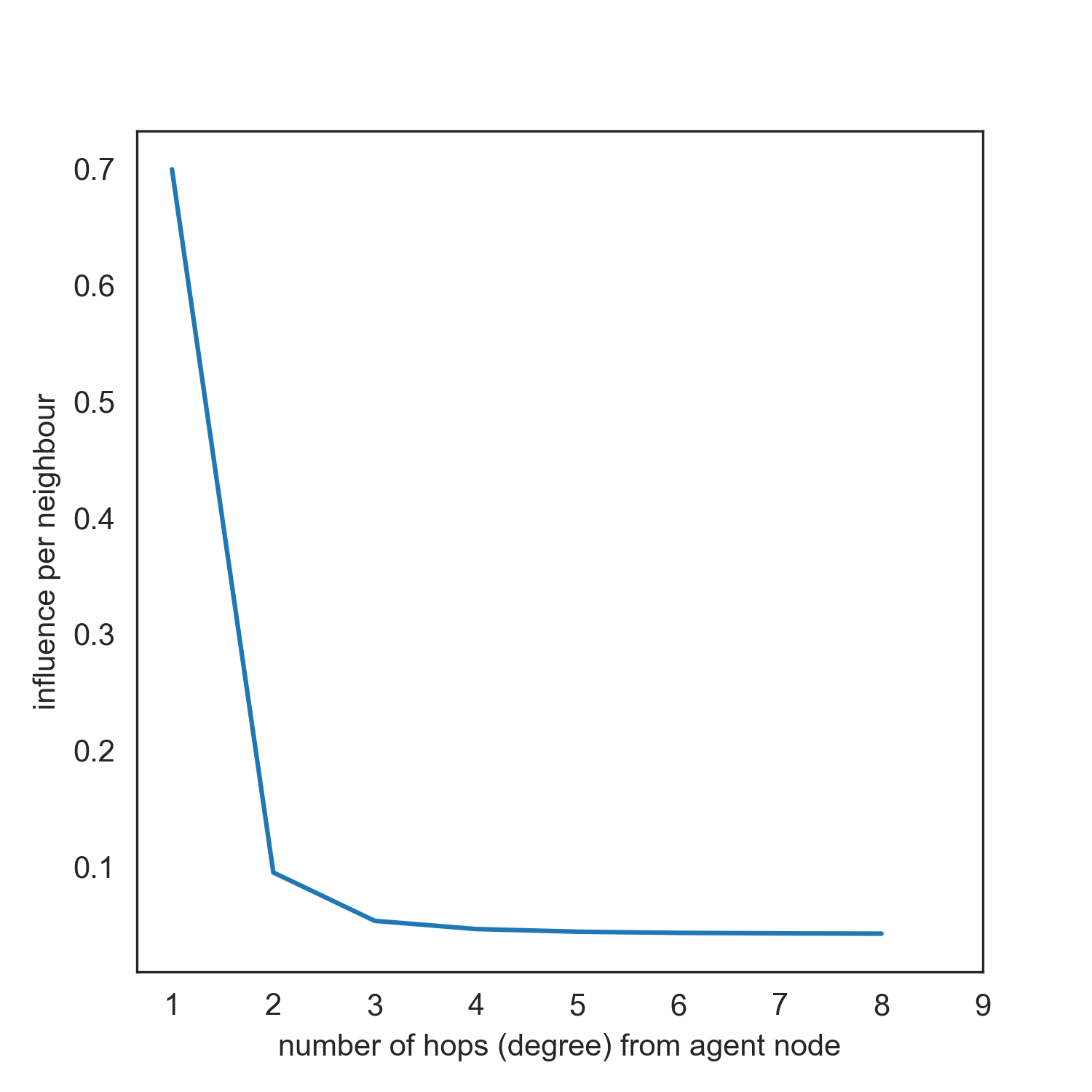}\label{fig:neighbourplot}}
  \caption{Illustration of influence weights calculation and neighbor plots}
\end{figure*}

Equation \ref{eq:baseinfluencegen} represents the base influence of stance upon an agent by the agent's neighbors. $I$ is an agent's influence stance outcome score, comprising the sum of stances of the neighbors in the agent's network. A first degree neighbor is a node that is one hop away from the agent, a second degree neighbor a node that is two hops away from the agent, and so forth. Based on the number of hops away from the agent, the influence of the node's stance on the agent decreases by a scalar multiple such that each neighbor in that hop contributes an equal influence on the agent in focus. This concept is borrowed from Katz Centrality. For the 1st degree neighbor, each neighbor $i$ of the total $n$ 1st-degree neighbors contributes $\frac{1}{n}$ influence on the agent; this is further reduced by a scalar multiple of $\frac{1}{m}$ for $m$ second degree neighbors and so on. 

\begin{equation}
\begin{split}
    I =  \frac{1}{n}\sum^n_{i=0}Y_\textnormal{{1st deg neighbors}} + \frac{1}{n}\frac{1}{m}\sum^n_{i=0}\sum^m_{j=0}Y_\textnormal{{2nd deg neighbors}} + \\
    \frac{1}{n}\frac{1}{m}\frac{1}{l}\sum^n_{i=0}\sum^m_{j=0}\sum^l_{k=0}Y_\textnormal{{3rd deg neighbors}} + ... 
\label{eq:baseinfluencegen}
\end{split}
\end{equation}

Figure \ref{fig:influenceweight} illustrates how neighbors are weighted based on their distance to the agent in focus for neighbors up to the 2nd degree. The influence weights $w$ of each neighbor is a function of the number of neighbors the nodes have and the distance to the agent.

We calculated the influence weights of each neighbor as the number of hops from the agent increases for 20,000 agents. The results are plotted in \ref{fig:neighbourplot}, in which by the elbow rule, the optimal number of hops away from an agent node is 2 hops. The influence per neighbor exponentially decays and tends to 0 as the number of hops increases. As such, our stance flipping prediction model considers only the influence of the first and second degree neighbors. The model that considers only the first degree neighbor influence is the Base Model, presented in Equation \ref{eq:baseinfluence}. Equation \ref{eq:seconddegneigh} extends the base model to evaluate the effect of adding the influence of second degree neighbors. We enhance this base model by adding mechanisms: stance strength, connection and reciprocity. 

\begin{equation}
\begin{split}
    I = \alpha\left[ \sum^n_{i=0}Y_\textnormal{{1st deg neighbors}} \right] \textnormal  {,where } \alpha=\frac{1}{n}
\end{split}
\label{eq:baseinfluence}
\end{equation}
 
\begin{equation}
\begin{split}
    I = \alpha\left[\sum^n_{i=0}Y_\textnormal{{1st deg neighbors}} + \sum^n_{i=0}\sum^m_{j=0}\beta Y_\textnormal{{2nd deg neighbors}}\right] \\
    \textnormal{ ,where  } \alpha=\frac{1}{n}, \beta=\frac{1}{m}
\end{split}
\label{eq:seconddegneigh}
\end{equation}

\textbf{Stance Strength.} The first mechanism we add is the effect of stance strength on an agent's outcome score: $Y_{agent} = \gamma X_*B_*$, where $\gamma$ is a scalar representing the agent's stance strength and its importance. Stance strength alludes to the fact that the more an agent expresses a stance, the stronger the belief in the stance. It is defined as the proportion of the final stance $s_{final}$ is expressed against the number of expressed stances $s$, multiplied by the variable importance value $w_s$. With this mechanism, neighbor's stances $Y$ are calculated similarly.
\begin{equation}
  \gamma = \frac{|s_{final}|}{|s|} \times w_s
\label{eq:stancestrength}
\end{equation}

\textbf{Connection.}
Connection $C$ is the proportion of neighbors that support an agent's stance:
\begin{equation}
    C_{agent} = \frac{\textnormal{\#neighbors with same stance}}{\textnormal{\#neighbors}}
\label{eq:connection}
\end{equation}

Connection represents opinion similarity between the agent and the agent's neighbors, which lends strength to the stance an agent expresses. Collectively, agent (and neighboring agent) stances are enhanced with this mechanism: $Y_{agent} = \gamma CXB$

\begin{figure*}[!tbp]
  \centering
  \includegraphics[width=1.0\textwidth]{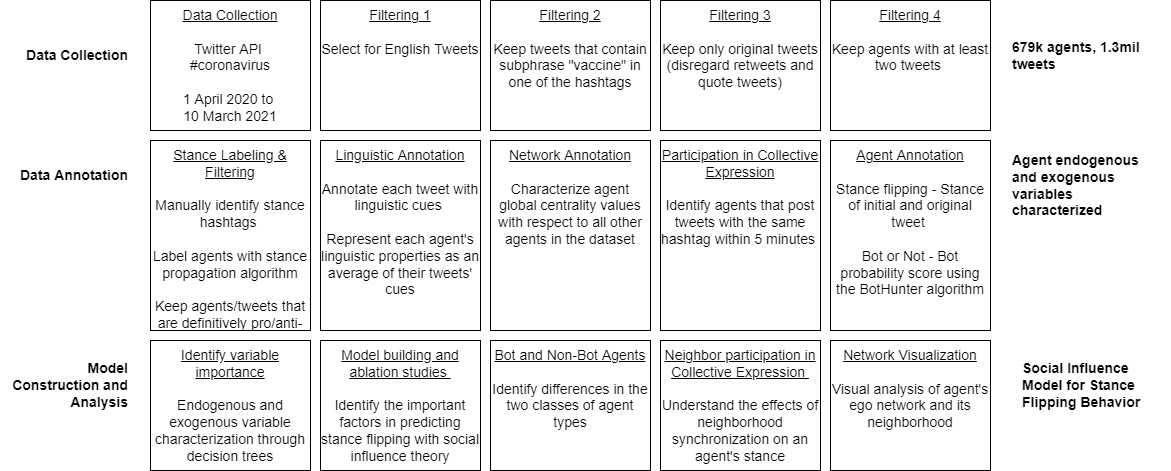}
\caption{Diagram for methodology of building the Social Influence Model for prediction of stance flipping behavior.}
\label{fig:methodology}
\end{figure*}

\textbf{Reciprocity.}
Reciprocity $R$ is the two-way interaction between two agents. The higher the reciprocity value, the closer the agents are in a friendship, leading to a higher influence on the agent.
\begin{equation}
    R = 2 \times \textnormal{\#reciprocal interactions}
\label{eq:reciprocity}
\end{equation}

This mechanism thus modifies the stance score of a neighbor agent: $Y_{neighbor} = \gamma CXB + R$
The stance score of the agent in focus remains the same.
    
\textbf{Susceptibility Score.} We define a susceptibility score $S = (I - Y_{agent})^2$ , which characterizes the difference in the score between the agent's stance and the influences from the variables. The higher the susceptibility score of agent $i$, the more likely the agent will flip its stance due to social influence. The agent $i$ will flip stance if $S_i\geq \epsilon$. For the base model, we set $\epsilon$ at 10\% of the number of agents. For the models with second degree neighbor information, $\epsilon$ is 1\% of the total number of agents, reflecting the proportion of the number of agents that flip stances in the overall dataset.


\subsection{Determining variable importance}
We need to determine the coefficients $B_*$ of the variables in the model. As such, we performed a binary classification task with a decision tree model using the Python sklearn library. We run this decision tree across the entire dataset to collectively determine feature importances.
The task uses all the defined linguistic and network variables to classify whether the agent flips or not. We performed a five-fold cross-validation with an 80-20 train-test split for each fold. To account for the huge class imbalance, we used the stratified sampling method, which makes sure that both the train and test sets have an equal proportion of both types of agents. We then extracted the feature importance from the decision tree model, and these feature importances are then used as the variable importance matrix $B_*$.

\subsection{Experimental setup}
We apply the social influence model to our dataset to predict stance flips. We only investigate agents who have more than one tweet in order to have changes in vaccine stances. 
For these agents, we leave out each agent's last stance, and use the collected historical data to predict the final stance. However, in the collected historical data, we do include agents that have only one tweet, as they contribute influence to the agents in focus. We progressively add mechanisms to the base model, studying the effects the effects of each variable on model performance. We measure the macro-F1 accuracy to factor for the unbalanced dataset. 
Then, we analyze the social influence and tendency to flip by the two classes of data: bots and non-bots.

\section{Data and Methodology}
Figure \ref{fig:methodology} describes the methodology of building the Social Influence Model for prediction of stance flipping behavior. In the following subsections, we detail each of the modules in the Data Collection and Data Annotation sequence. We described the Model Construction in Section \ref{sec:model} and present the experimental results and analysis in Section \ref{sec:results}.

\subsection{Data Collection}
We collected Twitter data surrounding the COVID pandemic using the Twitter V1 REST API using the hashtag \#coronavirus daily from 1 April 2020 to 10 March 2021, selecting only for English tweets. We began data collection after the Pfizer-Biotech vaccine began development (in March 2020). We filtered the data to tweets about vaccines, keeping the tweets that have the sub-phrase ``vaccine" in one of the hashtag.
Additionally, as we are specifically looking for agents that flip stances, we disregard agents that only have one tweet in the dataset. We also analyze only original tweets, removing retweets and quote tweets that do not represent a stance originating from the agent.
Finally, we have 679,235 agents and more than 1.3 million tweets.

\subsection{Data Annotation for the Social Influence Model}
To construct the social influence model, we first labelled the tweets with their stance and linguistic cues from the text. After this enhanced description of the tweets, we combined all the tweets posted by an agent to label agents in terms of their overall stance, network centrality, and mean linguistic cues. Finally, we use these information as input into our social influence model.

\textbf{Stance Labelling.} We manually inspected all vaccine-related hashtags in the dataset and classified the hashtags into pro- and anti-vaccine hashtags. We left out generic hashtags like ``\#vaccine" and ``\#covidvaccine" that do not present a stance. As the coronavirus pandemic evolved, the hashtags and meaning of hashtags used on social media evolved along. Hence, we manually determined a list of seed pro-vaccine and anti-vaccine hashtags in intervals of three-months worth of data through looking at the data.
The list is in Appendix \ref{sec:stance-relatedhashtags} and the hashtags are split into the two-month timeframes.

We use a network-based stance propagation algorithm that models a user-hashtag bipartite graph and propagate the stance labels between the two parts, providing a label and a confidence value for all tweets and agents \cite{kumar2020social}. We further filter the tweets to those with a defined pro-/anti-stance and their authoring agents.

\textbf{Linguistic Annotation.} Language gives us an insight into an agent's thoughts and emotions \cite{pennebaker2003psychological}, and we infer these measures by characterizing linguistic cues. 
To characterize messages of both groups, we use Netmapper\footnote{http://netanomics.com/netmapper/} software to count the frequency of lexical categories, including abusive absolutist, positive and negative terms. This builds on psycholinguistic theory associating particular words and expressions with behavioral, cognitive and emotional states \cite{tausczik2010psychological}. 

For each Tweet, NetMapper returns quantitative counts of these linguistic cues in the text from a hand-curated multi-lingual lexicon. The these cues are: positive sentiment, negative sentiment, number of terms representing an identity (num identities term), number of pronouns used, 1st/2nd/3rd person pronouns, number of family terms, number of exclusive terms, number of abusive terms. NetMapper also counts punctuation terms such as the average word length and number of exclamation points used. Lastly, the NetMapper software determines a reading difficulty score using the formula from the Flesch-Kincaid reading difficulty score, which was standardized by the U.S. Military gauges the ease of readability of a text by the number of words and syllabus of the words. This is calculated by: reading difficulty = 0.39 $\times$ average words per sentence + 11.8 $\times$ average syllabus per word - 15.59.
We included the tweet count to an agent's endogenous variables to indicate how expressive the agent is.

\textbf{Network Annotation.} We measure the indication of how an agent is influenced by his neighbors by characterizing social network variables. We calculated an agent's global centrality values with respect to all the other agents in the collected dataset. The centrality values are: number of followers, eigenvector centrality, total degree centrality, betweenness centrality, super friends and super spreaders. 
These variables denote how connected and influential the agents are in the network.

\textbf{Agent Annotation.} Finally, as agents are the focus of the model, we annotate each agent with their corresponding stance, linguistic and network values. We labeled each agent's stance as the stance of the agent's final collected tweet. We kept a chronological history of each agent's stance, which we used in identifying agent stance flipping. We take the mean of each agent's tweets linguistic cues as the agent's overall linguistic cues. Agents' network values are annotated using the values generated from the network annotation, which represents each agent's global centrality value. 
In these aggregation analyses, we are accounting for the fluctuations of an agent's endogenous and exogenous variables, and making the assumption that the agent's writing style and social network influence is generally consistent.  

We then label the stance flipping behavior for each agent. An agent is considered to have flipped his stance if the original tweet and the final tweet has opposite stances. While agents may have change their stances multiple times within the sequence of tweets, this work focuses on how prone agents are to flipping their stances. In doing so, we use only the observation that the agents had flipped their stance at least once, i.e. the difference of stances between initial and final stances, and not the frequency of stance flipping.

\subsection{Understanding agent and its neighborhood}
An important part of the social influence model is the agent's profile and neighborhood. 
To analyze the effect of an agent's profile with the social influence model, we identify each agent as a bot or not, by annotating each agent in terms of its bot-probability. 
To analyze an agent's neighborhood, we annotate each agent in terms of the agent's engagement in collective expression by borrowing ideas from coordinated action detection.

\textbf{Bot Annotation and Analysis.} We performed bot-probability annotation using the BotHunter algorithm at the 0.70 threshold level \cite{beskow2018bot}. The algorithm extracts account-level metadata and classifies agents using a supervised random forest method through a multi-tiered approach, each tier making use of more features. 
For each user agent, BotHunter provided a probability that the account is inorganic. A probability over 70\% indicates the agent is likely to be a bot \cite{ngstability}.
We also annotate users that self-identify as bots through having the word ``bot" in their username, i.e. ``coronaupdatebot".
We then analyze the differences in susceptibility scores between the bot and non-bot populations and visualize the network interaction structure.

\textbf{Engagement in collective expression.}
Our social influence model is centered around the agent itself.
To gain insight into the difference in the neighborhood between the agents that flip stances and those that do not, we investigated the neighborhood of both types of agents in terms of whether their neighbors participated in the collective expression of ideas. We borrow ideas from studies of detection of coordinated action and measure this through the extent of hashtag coordination between agents. To do so, we utilized the Synchronized Action Framework \cite{DBLP:journals/corr/abs-2105-07454}, which seeks to identify coordination between agents through the anomalously high volume of activity of synchronized action. 

In this study, we apply the framework to annotate agents as engaging in collective expression or not engaging in collective expression.
We define collective expression between two agents in terms of anomalously high numbers of tweets with the same hashtag within a short time window.
To measure collective expression, we first measure the number of times the agents post a tweet with the same hashtag within a 5-minute sliding window, constructing tuples of (agent1, agent2, number of similarities). 
We then determine the mean and standard deviation of the distribution of the number of similarities. This value is thus called the agent-pair similarity value.
We then filter the tuple dataset to retain only agent pairs that have an agent-pair similarity value that is greater or equal to (mean + 1 standard deviation). This step not only serves to reduce the data size but also serves to remove spontaneous expression where two agents coincidentally used the same hashtag within a short time period. 
We annotate this set of agents as engaging in collective expression, given that they have expressed similar opinions multiple times within a short time window. 
We overlay this information on the original set of agents. We then analyze the agents that flip their stances and those that do not in terms of the proportion of 2-degree neighborhoods participating in collective expression through the same hashtags. We also identify the proportion of 2-degree neighborhoods that have an opposite stance from the agent and participate in collective expression.

\section{Results}
\label{sec:results}

\subsection{Summary of Data}
Across the dataset, we collected 679k agents with 1.3 million tweets. The vaccine-related tweets were mostly in the languages English, French and Spanish. 32\% of the dataset are classified as bots and 1.6\% of the agents self-identify as bots.
Table \ref{tab:datasummary} summarizes the statistics of the dataset.

The proportion of stances for tweets and agents in the dataset are around the same: 90\% pro-vaccine and 10\% anti-vaccine. In total, only 1\% of the agents exhibited the stance flipping behavior. 
Although this work focuses only on the change in initial and final stances as the stance flipping behavior, we note that there are agents that change their stances more than once. About 0.4\% of users changed their stances more than once, with the highest percentage at three changes (0.2\%). The largest number of stance changes observed in our dataset is 30 stance changes over 485 tweets.

\input{tables/datasummary}

A larger percentage of agents flipped from pro-vaccine to anti-vaccine (61.2\%) compared to anti-vaccine to pro-vaccine (38.8\%).
Table \ref{tab:stanceflipexamples} shows two examples of agents that flipped from pro-vaccine to anti-vaccine stances. These are original messages, i.e. the messages are written by the agents themselves and are not retweets or quoted tweets.

A manual inspection of the tweets of agents before and after flipping stances reveals how providing vaccination to the masses can change their perspectives. Agents commonly move from the \#VaccineHesitancy rhetoric to the \#TakeTheVaccine promotion after taking their first dose of the Pfzier vaccine. These agents also further the narrative of \#EndTheLockdown at the same time, showing their eagerness to end the lockdown state governments have imposed to contain the virus spread. On the other hand, agents change their tune from \#VaccineSavesLives to \#VaccineDeath when they recount stories of loved ones having severe allergic reactions to the vaccine or beliefs in rumors about diseases that vaccines may cause (e.g. autoimmune disease). Another common move to the anti-vaccination camp is to stand for \#VaccineDiplomacy and \#VaccineEquality, in which all nations, regardless of diplomatic ties with other nations, and all people, regardless of socioeconomic status, have an opportunity to take the vaccine. When this fails, the solution offered is for everyone not to take the vaccine for equality.


\begin{table*}[!ht]
\begin{center}
\begin{tabular}{|p{7cm}|p{7cm}|}
\hline
\textbf{Agent 1} (pro- to anti-vaccine) & \textbf{Agent 2} (anti- to pro-vaccine)\\ \hline
When a business has a 20 times return on investments u push for it the best u can \#business \#VaccinesWork \#covid19  & \textit{[...] I will never have the vaccine ever, stop me going places, stop me working, I am me and that's enough \#NoVaccine \#COVIDVaccine} \\ \hline
\textit{\#CovidVaccine \#VaccinePassport \#COVIDIOTS \#covid19 why risk your precious health on a trial vacc for a disease with over 97\% recovery \& higher for young people} & March 2020 was the hardest. March 2021 slightly better due to \#vaccines\\ \hline 
\textit{Rachel's family have reported her death as a 'yellow card' as she developed her symptoms after receiving a coronavirus vaccine [...]} & Protect your community. \#getthevaccine\\ \hline
\end{tabular}
\caption{Original tweets showing agent stance flips. Standard text are pro-vaccine texts, \textit{italicized} text are anti-vaccine texts.}
\label{tab:stanceflipexamples}
\end{center}
\end{table*}

\subsection{Social Influence on COVID Vaccine Stance}
Based on a five-fold cross-validation run on a decision tree, the most important features are: (a) linguistic variables: number of tweets, average word and sentence length, reading difficulty;  (b) network variables: number of followers, eigenvector centrality, super spreaders and betweenness centrality. These importance values are used as the coefficients in $B_*$ in the social influence model. The importance scores of each endogenous and exogenous variables of the agents are reflected in the Appendix at Table \ref{tab:featureimportances}. When we compare the feature importance across all agents, bot agents only and non-bot agents, we observe some differences. Variables that are more important to bot agents compared to non-bot agents are: tweet count, positive/negative sentiment, 1st person pronouns, number of family/exclusive terms and eigenvector centrality; while super spreaders and super friends are less important variables to bot agents. Variables that are more important to non-bot agents are: number of pronouns used, 1st/2nd/3rd person pronouns, super friends and super spreaders; while eigenvector centrality, positive/negative sentiment and family/abusive terms are less important to non-bot agents.

\input{tables/results}

We perform incremental experimental runs on our dataset, each run adding a mechanism to the model. The results are presented in Table \ref{tab:modelresults}. Our final stance flipping model outperforms all the other models with an accuracy score of 86\%, showing that a combination of all the identified factors is important to the influence of the agent stances. Statistical tests of the model prediction results identified that factors important to stance flipping prediction are: the use of 2nd-degree neighbor information, stance strength and connection information, while the reciprocity factor does not achieve a significant improvement in results.

Ablation analysis where we removed either network or linguistic variables in the base model shows a low prediction score of around 0.17\% and 0.19\%, indicating that both linguistic and network variables contribute to the success of the model prediction. In the ablation analysis detailed in Table \ref{tab:modelresults}, we observe that the base social influence model performs significantly better than the same model without network or linguistic variables. In addition, the final model shows a significant difference in predictive ability between the bot and non-bot agents, suggesting that these two classes exhibit different stance flipping behavior.

\begin{figure*}[!tbp]
  \centering
  \subfloat[\textbf{Pro to anti} vaccine stance flip]{\includegraphics[width=0.4\textwidth]{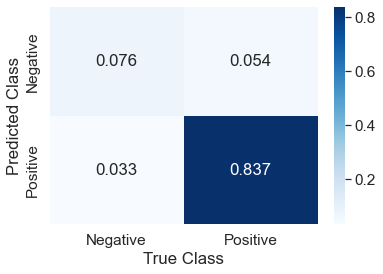}}
  \subfloat[\textbf{Anti to pro} vaccine stance flip]{\includegraphics[width=0.4\textwidth]{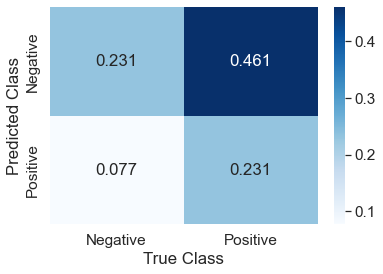}} \\
  \subfloat[\textbf{Bots} stances]{\includegraphics[width=0.4\textwidth]{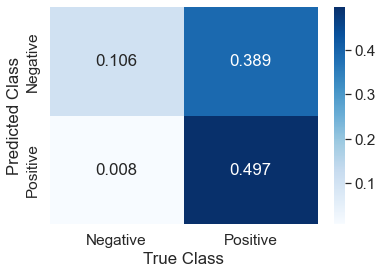}}
  \subfloat[\textbf{Non-Bots} stances]{\includegraphics[width=0.4\textwidth]{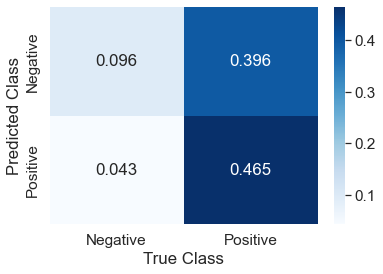}}
\caption{Confusion Matrices for prediction of stance flipping behavior by agent types, i.e. stance flip direction and bot class. These results are generated with the final model, Model 5.}
\label{fig:confusionmatrices}
\end{figure*}

Our baseline decision tree model performs at 37\% accuracy. Our base prediction model that takes in only the first degree neighbor information performs similarly to the decision tree model. The accuracy increases 11\% with the addition of information from second degree neighbors, indicating the importance of indirect influence on an agent's stance. While there is a slight 5\% increase in accuracy with the addition of connection, the accuracy again increases drastically at 11\% with the addition of reciprocal ties, showing that the stronger the tie between two agents, the stronger the influence mechanism.  
Investigating the confusion matrices of the final Model 5 (Figure \ref{fig:confusionmatrices}) we observe that the model best predicts true positives for a pro-to-anti stance flip and true negatives for an anti-to-pro stance flip.

Performing a two-sample t-test shows a significant difference in the means of the susceptibility scores of the agents that flip stances against the agents that do not ($t=2.97, p=0.003$). This test is performed at the 95\% confidence interval. Hence, $p=0.003<0.05$ indicates a separation of susceptibility scores between the two groups, with the agents that flip reflecting a higher susceptibility score. 

\subsection{Bot and Non-bot Agents}
Out of the agents that flip their stances, 53.7\% are identified as bots by the BotHunter algorithm. 6.6\% of the overall bot population flip stances, while only 2.7\% of non-bot agents flip stances. Bots are easier to predict, resulting in a higher accuracy score than non-bots agents. 
We show an example of an identified bot agent that flip stance in Table \ref{tab:botmessageexample}. This account repeats a message from the anti-vaccine camp several times before repeating a message from the pro-vaccine camp.

\input{tables/samplemessages}

\begin{figure}[!tbp]
\includegraphics[width=0.5\textwidth,height=0.2\textheight]{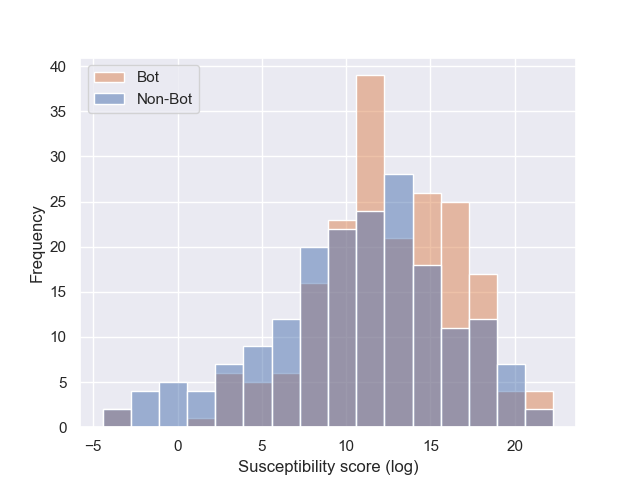}
\caption{Histogram of susceptibility scores (log scale) for bot and non-bot agents}
\label{fig:botdeflection}
\end{figure}

The bot population has a higher susceptibility score than the non-bot population and the overall population average, which is visualized in Figure \ref{fig:botdeflection}. The histogram of susceptibility scores of non-bots is shifted to the left, showing they generally form lesser interactions with other agents (connections/ reciprocal) and are more convicted on their stance.

\begin{figure*}[!tbp]
  \centering
  {\includegraphics[width=1.0\textwidth,height=0.3\textwidth]{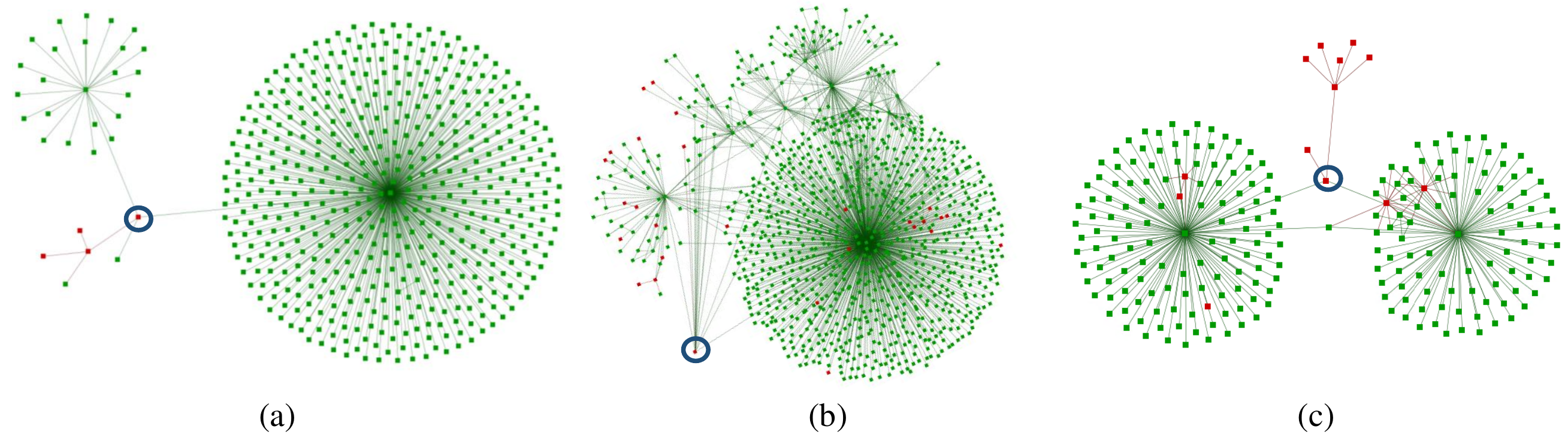}}
  \caption{Network interaction graphs of \textbf{bot agents} that our model correctly predicts to flip. Green and red nodes are pro- and anti-vaccine agents respectively. The color of the agent stance is the stance before the flip. Agents' usernames are redacted to maintain user privacy.}
  \label{fig:botgraphs}
\end{figure*}

\begin{figure*}[!tbp]
  \centering
  \includegraphics[width=0.9\textwidth,height=0.8\textwidth]{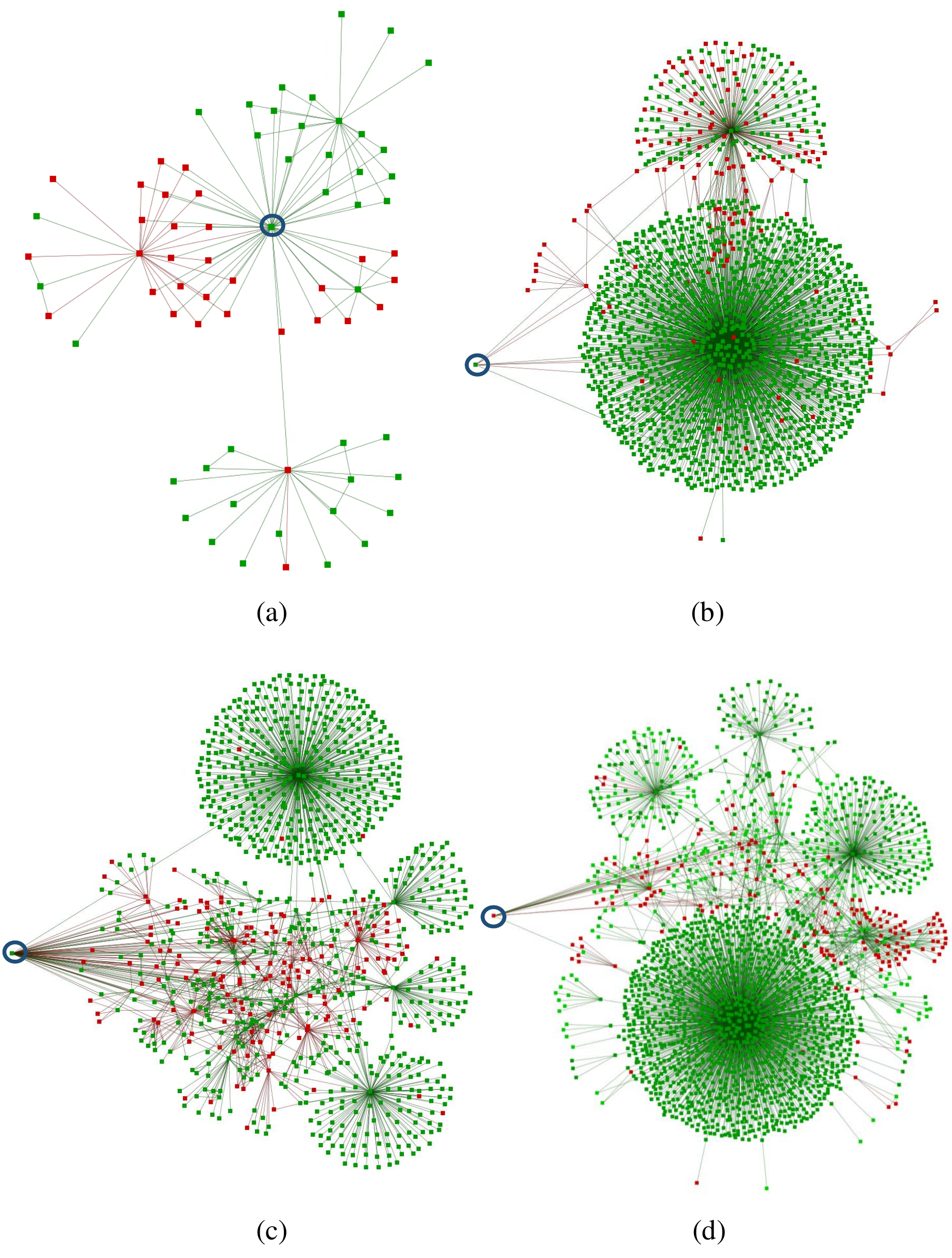}
  \caption{Network interaction graphs of \textbf{non-bot agents} that our model correctly predicts to flip. Green and red nodes are pro- and anti-vaccine agents respectively. The color of the agent stance is the stance before the flip. Agents' usernames are redacted to maintain user privacy.}
  \label{fig:resultgraphs}
\end{figure*}

Figure \ref{fig:botgraphs} and \ref{fig:resultgraphs} show positive results for bot and non-bot agents respectively. These network graphs represent the interaction between agents: a node is an agent and agents are linked if they interacted through a tweet mention, quote or retweet. Agents that are predicted to flip according to the social influence model and their final stance is indeed a flipped stance. In general, we observe that agents that are detected to flip have a very strong network influence of the opposite stance, emphasizing the importance of peer effects, where connected agents have a strong influence over an agent's opinion. Compared to non-bot agents that flip stances (Figure \ref{fig:resultgraphs}), non-bot graphs are typically very sparse and connected to one or two other large clusters.
Self-declared bot accounts with the word ``bot" in their user names typically have large susceptibility scores that are in the 95th-percentile zone of the susceptibility scores dataset. 5\% of these agents flip stances. 
Figure \ref{fig:confusionmatrices} shows the confusion matrices for the prediction of stance flipping behavior separated into bots and non-bot classes. The evaluation metrics are similar, showing that the model performs similarly on both agent types.

\begin{figure*}[!tbp]
  \centering
  \subfloat[\textbf{Bots} stances]{\includegraphics[width=0.5\textwidth,height=0.4\textheight]{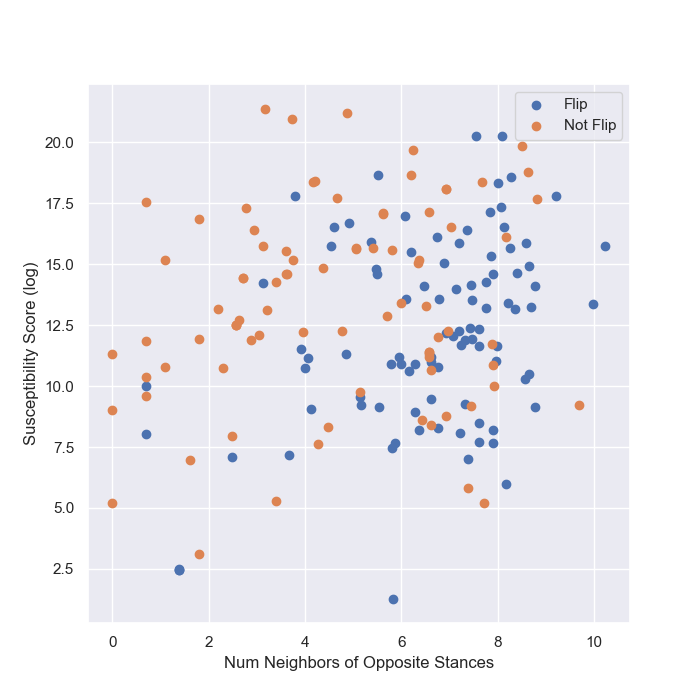}\label{fig:botgraph}}
  \subfloat[\textbf{Non-Bots} stances]{\includegraphics[width=0.5\textwidth,height=0.4\textheight]{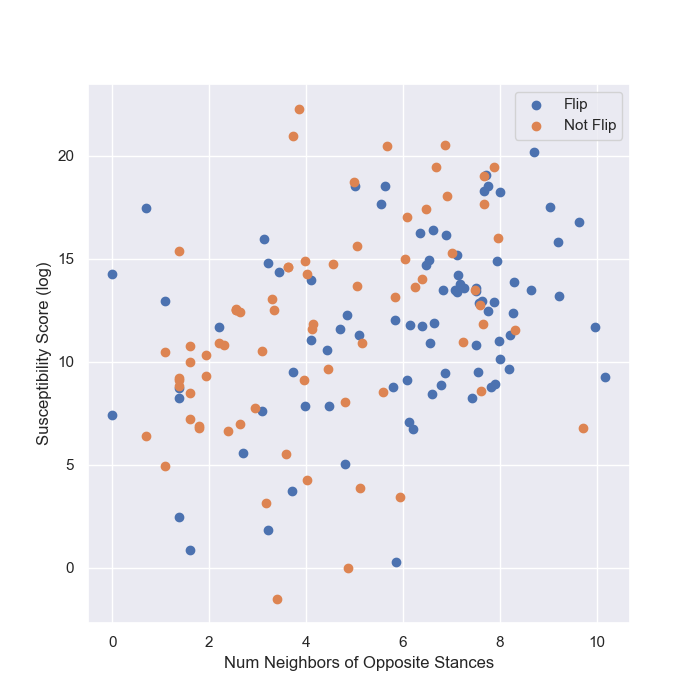}\label{fig:nonbotgraph}}
\caption{Susceptibility score plots of Bots and Non-Bot agents}
\end{figure*}

Figures \ref{fig:botgraph} and \ref{fig:nonbotgraph} depict the susceptibility scores against the number of 1st and 2nd degree neighbors that have the opposite stance. Bot agents have a good split of susceptibility scores: bot agents that flip stances have higher susceptibility scores, while agents that do not flip stances have lower susceptibility scores. They flip stances at a lower value of their neighbors' opposite stances than non-bot agents compared to non-bot agents, i.e. the number of first and second degree neighbors with stances opposite to the agent's current stance. 
Non-bot agents are harder to predict as there is no clear distinction between the susceptibility scores of agents that flip and those that do not. 

\subsection{Neighbors Engaging in Collective Expression}
We identified 0.1\%, or 6791 agents, engaged in collective expression. The minimum number of times an agent in this group posts a tweet with a hashtag that matches the hashtag of another agent in this group is 80.54 times, and the maximum number of times is 1704 times. 

\input{tables/coordination}

We perform a statistical t-test between the neighborhood of agents that flip stances and agents that do not and present the results in Table \ref{tab:coordination}. These statistical tests identify whether there are significant differences in the means between these two groups of agents. We note in both groups of agents, the type of agents in their neighborhood are similar, i.e. bot or non-bot. However, the participation of neighbors in collective expression is important to the stance flipping observation. By participation of collective expression, it means that that agents share hashtags. The more an agent's neighbors participate in collective expression, the more the neighbors seem to agree with each other through using the same hashtags. When these neighbors are of the opposite stance as the agent, the more likely the agent is to flip stances.

\begin{figure*}[!tbp]
  \centering
  {\includegraphics[width=0.5\textwidth]{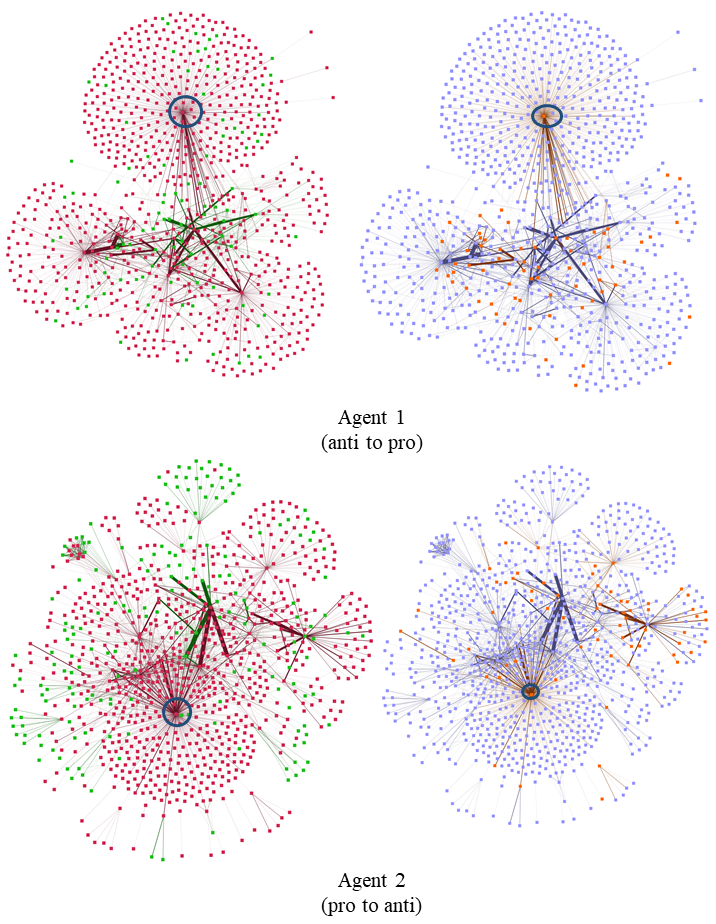}}
  \caption{Network interaction graphs that our model correctly predicts to flip. Green nodes are pro-vaccine agents; red nodes are anti-vaccine agents; orange nodes are agents found participating in collective expression through hashtags; purple nodes are agents that are not found participating in collective expression. The agent is circled in blue and the color of the agent stance is the stance before the flip.}
  \label{fig:combinedflip}
\end{figure*}

\begin{figure*}[!tbp]
  \centering
  {\includegraphics[width=0.5\textwidth]{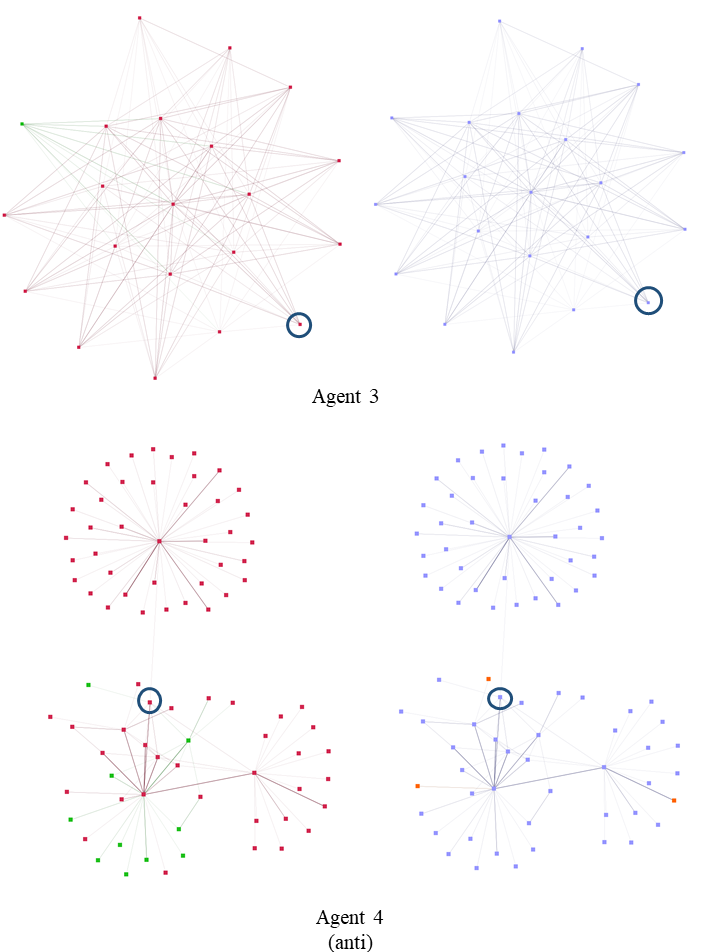}}
  \caption{Network interaction graphs that our model correctly predicts to not flip. Green nodes are pro-vaccine agents; red nodes are anti-vaccine agents; orange nodes are agents found participating in collective expression through hashtags; purple nodes are agents that are not found participating in collective expression. The two agents, circled in blue, are of the anti-vaccine stance.}
  \label{fig:combinednoflip}
\end{figure*}

Figures \ref{fig:combinedflip} and \ref{fig:combinednoflip} visualize the neighborhood of the agents that flip and do not flip stances respectively. These interaction graphs are constructed with agents as nodes and links as the number of interactions between agents through a tweet mention, quote or retweet. These graphs first show the stances of the agent and the neighborhood on the left, then show the agent's engagement in the collective expression of hashtags among the entire dataset.

Of the agents that flip stances, a significantly larger percentage are bots as compared to the agents that do not flip stances ($p=2.14e^{-13}<\alpha=0.05$).
This is consistent with our observation in the previous section that a larger proportion of the bot population flip stances.
In comparing the 2-degree neighborhood characteristics between both groups of agents, we observe that while there is no significant difference in the proportion of neighbors that are bots, agents that flip stances have a larger percentage of neighbors of the opposite stance.
This highlights that the expression of neighbor opinions (same or opposite stance) has a larger impact than the type of neighboring agents (bot or non-bot) on whether an agent flips his stance.

In terms of neighbors participating in the collective expression of opinions through the same hashtags within a short time period, agents that flip stances have a significantly higher proportion of such neighbors compared to agents that do not flip stances ($p=0.27<\alpha=0.05$). The group of agents that flip stances also have a higher proportion of neighbors that are both participating in collective expression and are of the opposite stance of the agent, signifying the peer pressure of an agent's neighborhood on the agent's expression of opinion.

Observations from the network visualizations of agent interaction show that agents that are predicted correctly to flip their stances have a denser neighborhood (i.e. more neighbors up to 2-degree away) than agents that do not flip their stances. These visualizations also show the higher proportion of neighbors engaging in collective expression surrounding agents that flip stances.

\section{Discussion}
This paper examined the coronavirus vaccine stance flipping phenomenon on Twitter. This phenomenon is rare, occurring in only 1\% of the agents we collected, indicating that most agents on Twitter do not change their stance once expressed publicly. While it is reassuring that there is a larger proportion of pro-vaccine to anti-vaccine tweets and agents collected, it is worrying that a larger proportion of agents flip their stance from pro-vaccine to anti-vaccine. This indicates the slight success of the anti-vaccination movement. 

Using a combination of the agent's endogenous and exogenous variables, we constructed a social influence model to predict whether an agent on Twitter will change stance toward the coronavirus vaccination. The model was incrementally built from the base influence model which estimates the impact of an agent's past tweets and the agent's neighbors on his stances, then additional mechanisms of stance strength, connection and reciprocity were added. We also further investigate whether social influence has differences between bot and non-bot accounts. 
In our estimate of linguistic variable importance, the variables word length, sentence length and Flesch–Kincaid reading difficulty score relates to the readability of the tweets. Tweets that are easier to read catches other agents attention better. 
We observe that more importance is placed on 2nd and 3rd person pronouns compared to 1st person pronouns. Pronouns highlight the attention of the author \cite{doi:10.1177/0261927X13502654}. 2nd person pronouns like ``you" directly address the reader and pull him closer to the author; 1st person pronouns like ``we" signifies the authors as embedded within a social relationship, making for a more inclusive conversation; 3rd person pronouns like ``she/he" expresses opinion of others as a distinct identity from the author. 
In our estimate of the network variable importance, we identify that eigenvector and betweenness centrality has high variable importance. These two measures signify the influence an agent has in a network, based on the concept of connections to influential agents and information flow respectively: an agent's position in the social network is one of the key factors in influencing others.

From the comparison of feature importances between bot and non-bot agents, for exogenous variables, bot accounts place more importance on eigenvector centrality while non-bot accounts place more emphasis on super friends/spreaders. This shows that bot accounts seek to influence others while non-bot accounts seek out influential agents in the network. In terms of endogenous variables, sentiment and types of terms (exclusive/family) are more important to bot accounts while pronoun types are more important to non-bot accounts, reflecting a personal vs automated interaction of bot and non-bot accounts respectively.

Our results contribute to the reflection of factors that influence an agent's stance in online social media: a combination of network and linguistic variables is crucial in predicting an agent's future stance. 
Social influence occurs through an individual's peripheral processing \cite{GASS2015348}.
An agent can be influenced by the opinions of the network of neighbors around him, as observed from the increased in accuracy after the addition of second degree neighbor information and reciprocal ties. The addition of incremental factors of our model shows that interaction information such as two degrees of neighbors and the connectivity of neighbors contribute significantly to the influence of an agent's stance. However, reciprocity does not achieve a significant improvement, suggesting that only a one-way interaction is required for social influence and two-way interactions are not necessary. In addition, one's conviction towards a stance plays an important role in the agent flipping behavior. In our model, this is represented by the significant improvement in model predictive ability when stance strength is added, an indication of how easily an agent can be influenced.

In comparing the stance flipping behavior of bots versus non-bot agents, not only do bot and non-bot agents exhibit significantly different stance flipping behavior, but we also observe that bot agents have lesser conviction and a larger proportion flip stances (6.6\%). Bot accounts flip even with fewer neighbors in the opposite direction of stances. In contrast, non-bot accounts have more conviction and a smaller proportion flip (2.7\%) flip stances and require more neighbors of the opposite stance to flip. For accounts that declare themselves as bot accounts by the use of the word ``bot" in their account name, the proportion of these agents flipping stances is five times higher than the population proportion. Agents in this group that flip stances tend to have large susceptibility scores, signifying that they mix with communities that are predominantly different from their original stance. 

Bot accounts typically repeat the same message from one stance several times before switching stances and repeating another message. While intuitively, we expect mis/disinformation bots to hold firm to their stance and not be impacted by influence from neighbors, we postulate that bot agents easily flip their stances to match their neighbors' stances to create the appearance of social cohesion. Social cohesion between communities can build echo chambers and increase polarization \cite{garcia2015ideological}. This involves further investigation from a longitudinal perspective. 

Bot agents are also easier to predict stance flips, as observed by the higher accuracy score than the non-bot group. This, together with bots requiring fewer neighbors of the opposite stance for a flipped stance, suggests that bots are more prone to flipping their stances. This opens avenue for further characterization and understanding of the neighborhood bots operate in and their behavior in opinion expression and changing opinion expressions.

Our observation that a significantly larger proportion of first and second degree neighbors participate in collective expression surrounding the agents that flip stances point to the contribution of social peer influence on an agent's expressed stance. A larger proportion of neighbors that are participating in this collective expression are also of the opposite stance. These neighbors constantly share tweets with the same hashtags temporally, filling up an agent's timeline and common-hashtag view on the agent's Twitter landing page with a disproportionate number of tweets and hashtags of the opposite stances. This collective expression by an agent's neighbors creates an illusion that the agent's peripheral group is all of the opposite stance, adding social pressure on the agent to flip stances. This could also be attributed to the cognitive bias called group think, where agents flip stances to achieve conformity within their neighborhood and reduce conflict.

Identifying collective expressions through hashtags is but only one way to characterize neighbor-to-neighbor interaction. Other explored ways of identifying collective expressions include the use of similar URLs and @-mentions within a short time frame \cite{DBLP:journals/corr/abs-2105-07454}. Studying these expressions may provide additional dimensions of how an agent's neighborhood affects an agent's opinion expression. 

Several limitations nuance our conclusions from this work. Users with extreme opinions are typically more vocal on social media, suggesting caution in extrapolating findings. The list of pro-vaccine and anti-vaccine hashtags needs to be continually updated as new hashtags emerge in social media lingo. To combat this, we have changed the hashtags used every three months based on initial glimpses of the data. However, using singular hashtags can be insufficient in determining the stance of a Tweet. For example, the anti-vaccine label \#FakeVaccine could also be inserted into a Tweet about a story of receiving water instead of the actual vaccine. Determining stance labels through hashtags is thus a work in progress.

Next, the social influence model that we built predicts the vulnerability of an online agent to stance flipping. While it successfully predicts the stance flipping observations of our datasets, future directions involves expanding this model to incorporate temporal information such as the frequency of the stance flipping behavior.

Additionally, given our observations in this study that there are differences in the neighborhood between agents that flip stances and agents that do not, future work includes the expansion of our stance flipping model from focusing on an agent itself to factoring in neighbor-to-neighbor interaction information.

Lastly, given that the pandemic has evolved since this study began, future work could involve investigating users who were hesitant during the early pandemic period and their similarity/difference in stance after a period of time. The insights from these will aid in understanding key factors in changing an online user's stance, and hopefully inform future health information dissemination methods.

Nonetheless, we hope that our work provides an understanding of characterizing agents who flip their stances on Twitter, and provides an insight into the difference in influence bot and non-bot agents require to change their stance. In future work, we hope to incorporate apriori assumptions about content like an agent's personal values like political identity \cite{hornsey2020donald} in our stance flipping model.

\section{Conclusion} 
In this study, we observe stance changes towards the coronavirus vaccine on Twitter from April 2020 to May 2021, where 1\% of the agents exhibit the stance flipping behavior. 
To predict stance changes, we propose a novel model of calculating user susceptibility to stance dynamics in the Twitter social network which integrates linguistic information from an agent's past tweets and interpersonal influence from an agent's network connection. The model predicts whether an agent will flip stance with 86\% accuracy. 
Our results identify that agents that flip stances have significantly more neighbors engaging in a collective expression of the opposite stance, providing ideas that strong neighborhood expression can influence an agent's stance.
In a contrast analysis between bots and non-bot agents, we identify that a larger proportion of bot agents flip and flip even with fewer neighbors in the opposite direction of stances, signifying the social influence on these agents can be lesser compared to non-bot agents for them to change their stance.

\section{Acknowledgements}
The research for this paper was supported in part by the Knight Foundation, Cognizant Center of Excellence Content Moderation Research Program, the Office of Naval Research (grant Bothunter N000141812108), Scalable Technologies for Social Cybersecurity/ARMY (grant W911NF20D0002), CyberFit/Air Force Research Laboratory (grant FA86502126244) and by the center for Informed Democracy and Social-cybersecurity (IDeaS) and the Center for Computational Analysis of Social and Organizational  Systems (CASOS) at Carnegie Mellon University. The views and conclusions  are those of the authors and should not be interpreted as representing the official  policies, either expressed or implied, of the Knight Foundation, Office of Naval Research or the US Government.

\appendices
\input{tables/featureimportance}
\input{tables/stancelist}

\ifCLASSOPTIONcompsoc
  \section*{Acknowledgments}
\else
  \section*{Acknowledgment}
\fi

The research for this paper was supported in part by the Knight Foundation and the Office of Naval Research grant N000141812106 and by the center for Informed Democracy and Social-cybersecurity  (IDeaS) and the center for Computational Analysis of Social and Organizational  Systems (CASOS) at Carnegie Mellon University. The views and conclusions  are those of the authors and should not be interpreted as representing the official  policies, either expressed or implied, of the Knight Foundation, Office of Naval Research or the US Government.

\begin{IEEEbiography}
    [{\includegraphics[width=1in,height=1.25in,clip,keepaspectratio]{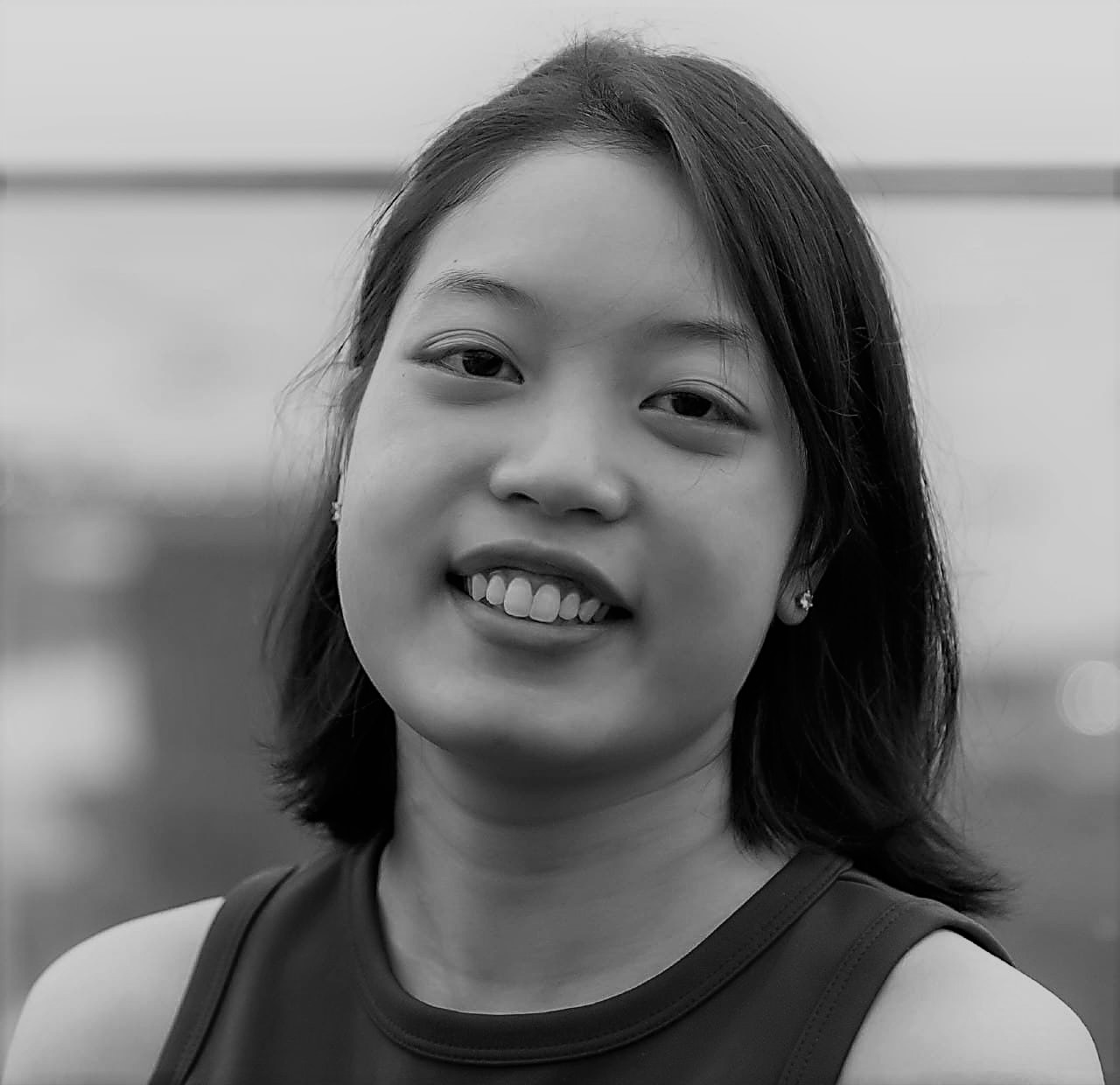}}]
    {Lynnette Hui Xian Ng}
is a PhD student in Societal Computing at Carnegie Mellon University. As a graduate researcher at the Center for Informed Democracy and Social Cybersecurity (IDeaS), her research examines social cybersecurity and digital disinformation. She holds an undergraduate degree in computer science from the National University of Singapore. Email: huixiann@andrew.cmu.edu
\end{IEEEbiography}

\begin{IEEEbiography}
    [{\includegraphics[width=1in,height=1.25in,clip,keepaspectratio]{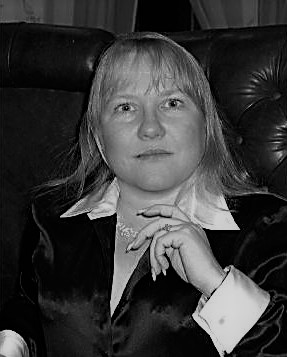}}]
{Kathleen M. Carley} is a Professor of Societal Computing, Institute for Software Research, Carnegie-Mellon University; Director of the Center for Computational Analysis of Social and Organizational Systems (CASOS), Director of the Center for Informed Democracy and Social Cybersecurity (IDeaS), and CEO of Netanomics. Her research blends computer science and social science to address complex real world issues such as social cybersecurity, disinformation, disease contagion, disaster response, and terrorism from a high dimensional network analytic, machine learning, and natural language processing perspective. She and her groups have developed network and simulation tools, such as ORA, that can assess network and social media data.  Email: carley@andrew.cmu.edu
\end{IEEEbiography}

\ifCLASSOPTIONcaptionsoff
  \newpage
\fi

\newpage
\bibliographystyle{IEEEtranN}
\bibliography{references}

\end{document}

%% file: tables/datasummary.tex
\begin{table}[!ht]
\begin{center}
\begin{tabular}{|p{4cm}|p{4cm}|}
\hline
\textbf{Criteria} & \textbf{Number of Agents}\\ \hline
Entire dataset & 679k \\ \hline
& \\ \hline
\multicolumn{2}{|l|}{\textbf{By Agent Type (Proportion)}} \\ \hline
Bot Agents & 32 \newline 1.6\% self-identify as bots\\ \hline 
Non-Bot Agents & 68\\ \hline 
\multicolumn{2}{|l|}{\textbf{By Final Stance (Proportion)}} \\ \hline
Pro-Vaccine Agents & 90\\ \hline 
Anti-Vaccine Agents & 10\\ \hline 
\multicolumn{2}{|l|}{\textbf{By Flipping Stance Observation (Proportion)}} \\ \hline
Agents that do flip stances & 1 \newline 0.4\% change their stance more than once\\ \hline 
Agents that do not flip stances & 99\\ \hline 
\end{tabular}
\caption{Summary of Data}
\label{tab:datasummary}
\end{center}
\end{table}

%% file: tables/results.tex
\begin{table*}[!ht]
\begin{center}
\begin{tabular}{|p{3cm}|p{8cm}|p{1.5cm}|}
\hline
\textbf{Model \#} & \textbf{Model} & \textbf{Accuracy} \\ \hline
Baseline & Decision Tree & 0.38 \\ \hline
Model 1 & Base social influence model  & 0.37 \\ \hline
Model 2 & Model 1 + 2nd deg neighbor information & 0.48* \\ \hline
Model 3 & Model 2 + stance strength & 0.70* \\ \hline 
Model 4 & Model 3 + connection & 0.75* \\ \hline 
Model 5 & Model 4 + reciprocity & 0.86 \\ \hline 
\multicolumn{3}{|l|}{\textbf{Ablations}} \\ \hline
Model 1 -  network & Base social influence model without network variables & 0.17* \\ \hline 
Model 1  - linguistic & Base social influence model without linguistic variables & 0.19* \\ \hline
Bots only & Model 5 with only bot agents  & 0.73* \\ \hline 
Non-Bots only & Model 5 with only non-bot agents & 0.67* \\ \hline 
\end{tabular}
\caption{Results of Social Influence Models. The base social influence model is defined in Equation \ref{eq:baseinfluence}. * indicates a significant difference at the $p<0.05$ level. For the models, the significant testing was performed against the previous model in sequence, and for the ablations, the significant testing was performed against Model 1.}
\label{tab:modelresults}
\end{center}
\end{table*}

%% file: tables/samplemessages.tex
\begin{table}[!ht]
\begin{center}
\begin{tabular}{|p{1cm}|p{7cm}|}
\hline
\textbf{Stance} & \textbf{Tweet} \\ \hline
antivax & DO NOT TAKE THE \#COVID VACCINE \#VaccineInjury \#VaccineDamage \#covidHOAX \newline \textit{message repeated several times} \\ \hline 
antivax & Mum passed away after taking experimental vaccine \#vaccinekills \newline \textit{message repeated several times} \\ \hline 
provax & ``Safe and effective" \#coronavirus \#Covid\_19 \#CovidVaccine \#VaccinesWork \#vaccinated \newline \textit{message repeated several times} \\ \hline 
provax & I was proud to get the COVID-19 vaccine earlier today at Morehouse School of Medicine.  I hope you do the same! \\ \hline
\end{tabular}
\caption{Sample messages of a bot agent that flip stances}
\label{tab:botmessageexample}
\end{center}
\end{table}

%% file: tables/coordination.tex
\begin{table*}[!ht]
\begin{center}
\begin{tabular}{|p{7cm}|p{3cm}|p{3.3cm}|p{1.5cm}|}
\hline
\textbf{Criteria} & \textbf{Agents that do flip stances} & \textbf{Agents that do not flip stances} & \textbf{p-value} \\ \hline 
\textbf{Proportion of bots} & 0.452 & 0.293 & 2.14e-13* \\ \hline 
\textbf{Proportion of neighbors that are bots} & 0.352 $\pm$ 0.372 & 0.358 $\pm$ 0.280 & 0.051 \\ \hline 
\textbf{Proportion of neighbors of the opposite stance} & 0.409$\pm$0.443 & 0.196$\pm$0.271 & 0.011* \\ \hline
\textbf{Proportion of neighbors participating in collective expression} & 0.0389$\pm$0.068 & 0.0302$\pm$0.115 & 0.027* \\ \hline
\textbf{Proportion of neighbors participating in collective expression and are of opposite stance} & 0.0207$\pm$0.0996 & 0.0136$\pm$0.0350 & 0.0078*\\ \hline
\end{tabular}
\caption{Comparison of agents that flip stances and do not flip stances.
Fraction of neighbors that are 1- of 2-degree away from these agents that meet various criteria are compared.} * denotes a p-value that is significant at the 0.05 significance level.
\label{tab:coordination}
\end{center}
\end{table*}

%% file: tables/featureimportance.tex
\section{Variable Importance}
Table \ref{tab:featureimportances} refers to the importance of endogenous and exogenous variables of the agents. 
\begin{table*}[!ht]
\begin{center}
\begin{tabular}{|p{4cm}|p{3cm}|p{3cm}|p{3cm}|}
\hline
\textbf{Variable} & \multicolumn{3}{|l|}{\textbf{Importance}} \\ \hline
& All Agents & Bot Agents only & Non-Bots Agents Only\\ \hline
\multicolumn{4}{|l|}{\textbf{Linguistic Variables}} \\ \hline 
Tweet count & 0.224 & 0.235 & 0.218 \\ \hline 
Avg word length & 0.068 & 0.067 & 0.069\\ \hline 
Reading difficulty & 0.058 & 0.065 & 0.063\\ \hline 
Positive sentiment & 0.040 & 0.050 & 0.036\\ \hline 
Negative sentiment & 0.029 & 0.034 & 0.031\\ \hline 
Agent's own stance & 0.026 & 0.025 & 0.027\\ \hline 
Num identities terms & 0.026 & 0.025 & 0.018\\ \hline
Num pronouns used & 0.021 & 0.022 & 0.021\\ \hline 
1st person pronouns & 0.015 & 0.019 & 0.022\\ \hline 
2nd person pronouns & 0.021 & 0.024 & 0.025\\ \hline 
3rd person pronouns & 0.018 & 0.021 & 0.026\\ \hline 
Num exclamation points & 0.015 & 0.018 & 0.020\\ \hline 
Num family terms & 0.009 & 0.018 & 0.006\\ \hline 
Num exclusive terms & 0.008 & 0.010 & 0.010\\ \hline 
Num abusive terms & 0.007 & 0.009 & 0.004\\ \hline 
\multicolumn{4}{|l|}{\textbf{Network Variables}} \\ \hline
Num of followers & 0.194 & 0.191 & 0.194\\ \hline 
Eigenvector centrality & 0.087 & 0.106 & 0.079\\ \hline 
Super spreaders & 0.043 & 0.019 & 0.057\\ \hline 
Betweenness centrality & 0.033 & 0.023 & 0.040\\ \hline 
Super friends & 0.022 & 0.019 & 0.034\\ \hline 
\end{tabular}
\caption{Variable importance from the decision tree model for all the agents in the dataset, and subsets of bot agents vs non-bot agents.}
\label{tab:featureimportances}
\vspace{-0.7cm}
\end{center}
\end{table*}


%% file: tables/stancelist.tex
\section{Stance-related hashtags}
\label{sec:stance-relatedhashtags}
The following hashtags were used to annotate pro- and anti- vaccine stances. The hashtags were selected based on manual examination of the unique hashtags in the data. Due to the fast changing nature of social media data, the hashtags were revisited every three months.

\small
\textbf{Pro-Vaccine hashtags:} 

\underline{April - Jun 2020:} VaccinesWork, Sharethevaccine, ProtectVaccineProgress, getvaccine, WaitforVaccine, FreeTheVaccine, vaccinesaresafe, vaccineconfidence, igotvaccinated, coronavaccineforall, CoronavirusVaccineAppointment, justtakethefuckingvaccine, VaccinesWithoutBorders, Vaccines4All, vaccinesaves, Vaccine4ALL, BreakthroughVaccine, waitingformyvaccine, GetTheVaccine, GetYourVaccine, takethevaccine, vaccinesafe, Vaccineswillwork, Iwilltakethevaccine, vaccinefreedom, VaccineToAll, SafeAndEffectiveCovid19Vaccine, VACCINESARESAFE, safevaccines, nosleeptilvaccine, NoOnsiteSchoolsUntilVACCINES, WhereIsMyVaccine, vaccineselfie, vaccineready, vaccinesareamazing, vaccinee, VirusToVaccine2020, HoHoHopeVaccineArrivesSoon, Vaccined

\underline{July - Sept 2020:} VaccinesSaveLives, havevaccinewilltravel, VaccinesWork, ArmysForVaccines, VaccinesForAll, VaccineForAll, We4Vaccine, vacciner, TakeTheVaccine, affordablevaccine4all, Votes4Vaccine, stopvaccinepolitics, SafeVaccines, NeedVaccine, GetTheVaccine, TrustYourVaccine, YesToCovid19Vaccine, WinWithVaccine, VaccineDay, VaccineSelfie, WheresTheVaccine, HaveTheCovidVaccine, coronavaccineforall, CovidVaccineToday, SecondVaccine, VaccineisSafe, vaccinesafetyadvocate, VACCINEROUNDTHECLOCKNOW, igotmycovid19vaccine, VaccineWorks, vaccineacceptance, VaccinesAreSafe, getmorevaccine, weneedthevaccine, TeamVaccines, TeamVaccine, GoForVaccine, getVaccine, provaccine, India4Vaccine, vaccinesmatter, VaccineNow, vaccinesavelivess, VaccineFTW, HaveTheVaccine, GetYourVaccine, vaccinesaveslives, VaccineWork, IWantMyVaccine, ProVaccines, vaccinesareamazing, makevaccinesfree, vaccine4all, rolloutthevaccine, WhereIsMyVaccine, VaccineSavesLives, TrustTheVaccine, covid19vaccineyesplease, getthecovidvaccine, covid19vaccine4all, SayYesToVaccine, ineedthevaccine, GetUsVaccines

\underline{Oct - Dec 2020:} VaccinesWork, VaccinesSaveLives, Vaccine4All, effectivevaccine, VaccineForSA, VaccinesSavesLives, SafeVaccines, accesstovaccines, VaccineHope, VaccinesWithoutBorders, wherearethevaccines, VaccinesforAll, getthevaccine, giveusthevaccine, VaccineWorks, vaccinesavelifes, GetAVaccine, vaccinesafelife, safevaccine, HaveTheVaccine, WhyIGotMyVaccine, CovidVaccineforall, vaccinesavestheworld, ImGettingTheVaccine, FreeVaccines, VaccinesWorkforAll, GiveMeMyVaccineNow, Affordablevaccine4all, getthatvaccine, justgivemethevaccine, SayYestotheVaccine, TakeYourVaccine, provaccine, YesToVaccine, vaccinesave, covid19\_vaccine\_4all, vaccineissafe, PleaseGetTheVaccine, VaccinesForEveryone, TrustTheVaccine

\underline{Jan - Mar 2021:} vaccinessavelives, VaccinesWork, Vaccine4All, TakeTheVaccine, FirstDoseOfVaccine, We4Vaccine, vaccinesaveslives, VACCINEFORWELLNESS, TheVaccineIsSafe, IWillTakeTheVaccine, firstdosevaccine, IWillTakeVaccine, SafeVaccine, SayYesToVaccine, VaccinesWorkForAll, VaccineIsSafe, TrustTheVaccine, getyourvaccine, SafeVaccines, VaccineSavesLives, vaccinesaresafe, YesToCovid19Vaccine

\textbf{Anti-Vaccine hashtags: }

\underline{April - Jun 2020:}NoVaccines, NoVaccinesForMe, VaccineYourAss, NoToCoronavirusVaccines, SayNoToVaccine, NoMandatoryVaccine, VaccinesKill, VaccinesKills, stopvaccine, fkyourvaccines, antivaccine, AntiVaccine, NoVaccine, StopCovidVaccine, WeDoNotConsentCVVaccine, VaccinesKill, VaccinesHarm, SayNoToVaccines, ForcedVaccines, VaccineIsPoison, ResistVaccines, Noneedvaccines, FakeCoronaVaccine, VaccinesAreBioweapons, Iwontgetthevaccine, VaccineFromHell, vaccinepoison, StopAllVaccines, Vaccinetakedown, vaccinesDAMANGEimmunity, vaccinesRnotNATURAL, RejectTheVaccines, BewareVaccines, FuckVaccines, HellNoVaccine, NoVaccinesEver, WhoNeedsVaccine, justsaynotoforcedvaccines, StickTheVaccineUpYourArse, nottakingavaccine, vaccinebad, WeaponizedDeadlyVaccines, JustSayNOToTheVaccines, DoNotTakeTheVaccine, FuckVaccinePoison, NOVaccine4Me, VaccinesNotSafe, VaccineBioWeapon, wedontwantvaccine, vaccinenotforme, DontTakeCovidVaccine, NotTakingCovidVaccine, DangerousVaccine, vaccinedoesnotwork, VaccineIsntSafe, No2Vaccine, OpposeTheVaccine, YouCanHaveMyVaccine, ShoveThatVaccineUpYourAsshole, MoThankYouCovidVaccine, ToHellWithCovidVaccine, CovidVaccinePoison, SCREWTHEVACCINE, murdervaccine, VaccineIsUseless, KilloronaVaccine, DodgyVaccine, DoNotTakeAnyVaccine

\underline{July - Sept 2020:} VaccineMortality, NoVaccine, NoVaccineForMe, NOVACCINE4ME, VaccineDeaths, VaccineDeath, VaccineInjury, AntiVaccine, vaccinesideeffect, NoToCoronaVirusVaccines, AvoidCovidVaccine, NoVaccines, coronavirusvaccinescam, ForcedVaccines, destroyvaccines, VaccineHesitancy, noneed4vaccine, notocovidvaccine, vaccinesharm, vaccinedamage, vaccinesareevil, SayNoToVaccines, killervaccine, no2vaccine, vaccinekills, vaccineskill, VaccinesKillingpeople, JustSayNotoVaccines, vaccinedanger, donttrustthecovid19vaccine, RejectWeaponizedVaccines, StopVaccine, FakeVaccine, vaccinechemicalweapon, DeathToVaccines, VaccineScam, NoVaccinesNeeded, NoVaccinesForMe, vaccinehesitant, PoisonVaccine, DeathVaccine, Cancervaccine, VaccineFail, vaccineRESISTANT, VaccineNonsense, UnsafeVaccines, NoVaccinesNecessary, AbnominableVACCINE, Vaccinefuckup, novaccinerequired, vaccinedrivemutations, VaccineFromHell, GodIsMyVaccine, AntiCovidVaccine, NeitherDoTheseCOVIDVaccines, notovaccine, Notmyvaccine, CovidVaccineIsPoison, VaccineDisaster, NovavaxVaccine, TheCovidVaccineKills, vaccinekills, StopTheVaccines, fuckyourvaccine, VaccineIssues, vaccinesDONTwork, vaccinebad, VaccineNotTheAnswer, dontgetthecovidvaccine, MurderbyVaccine, vaccineforwhat"

\underline{Oct - Dec 2020:} JeNeMeVaccineraiPas, NoVaccineForMe, stopthevaccine, novaccine, BoycottIndianVaccine, vaccinehesitancy, To\_Vaccine\_Is\_My\_Choice, vaccinefailure, donttakethevaccine, FakeVaccine, StolenVaccines, NoVaccine4Me, TrudeauVaccineContractsLie, vaccinesKill, VaccinesArePoison, VaccinesAreNotCures, NotAVaccine, jesusisvaccine, HALTtheVaccines, NoVaccines, dontgetthevaccine, VaccinesHarm, TeamNoVaccine, TheVaccineIsTheVirus, fakeVaccines, killervaccine, lethalvaccine, CoronaVaccineFail, vaccineBioweapon, saynotovaccines, Fuckvaccines, fuckyourvaccines, vaccinedeath, JustSayMoToVaccines, NoCOVIDVAccineMandate, NoMandatoryVaccines, GoHomeLeaveOurVaccinesAlone, antivaccine, anti\_vaccine, stopcovidvaccine, CovidVaccineHesitancy, VaccinesCanKill, Antivaccines, notovaccine, DontGetVaccine, VaccineDontWork, VaccineKills, StopVaccine, ForcedVaccine, FakeCovidVaccine, VaccinesAreBad, Falsevaccine, PfizerVaccineKillingPeople, NoCovidVaccineForMe, VaccineShaming, TakeTheVaccineAndShoveItUpYourAss, FakeVaccinesWillNotSaveYou, WorldSaysNoVaccine, NOCovidVaccine, NOCovid19Vaccine, Jesusovervaccines, IAmNOTVAccineBait, NotAVaccineAMedicalExperiment

\underline{Jan - Mar 2021:} norealvaccine, NoVaccine, VaccineScam, VaccineDeaths, NotAVaccine, VaccineKills, NOVACCINE4ME, vaccineisdeath, vaccineispoison, NoVaccineForMe, To\_Vaccine\_Is\_My\_Choice, vaccinesharm, NoVaccinePassports, NoToVaccinePassports, VaccinesAreDangerous, VaccinesKill, VaccinesAreBad, saynotovaccines, FuckTheVaccine, StillAintGettingTheVaccineThough, ChineseVaccineBioWeapon, antivaccine, JustSayNotoVaccines, Notomandatoryvaccines, fakevaccines, vaccinedeath , vaccineFail, StopVaccinePassports, CovidvaccineFail, COVID19VaccinesBioWeaponsOfMassDestruction, NoMandatoryVaccines, DoNotGetTheVaccine